\documentclass[reprint,superscriptaddress,amsmath,amssymb,aps,prl,floatfix]{revtex4-2}

\usepackage{graphicx}
\usepackage{tikz}
\usepackage{bm}
\usepackage{physics}
\usepackage{xcolor}
\usepackage[colorlinks=true,linkcolor=blue,urlcolor=blue,citecolor=blue]{hyperref}

\begin{document}

\title{Generalized Shift Vector as the Intrinsic Dipole \\ of Many-Body Correlated Electronic States}


\author{Jiaming Hu}
\email{hujiaming@westlake.edu.cn}
\affiliation{Department of Materials Science and Metallurgy, University of Cambridge, Cambridge CB3 0FS, United Kingdom}
\affiliation{School of Engineering, Westlake University, Hangzhou 310030, China}

\author{Sudipta Kundu}
\affiliation{Department of Materials Science and Metallurgy, University of Cambridge, Cambridge CB3 0FS, United Kingdom}

\author{Zhichao Guo}
\affiliation{Center for Quantum Matter, School of Physics, Zhejiang University, Hangzhou 310058, China.}

\author{Joshua J.P. Thompson}
\affiliation{Department of Materials Science and Metallurgy, University of Cambridge, Cambridge CB3 0FS, United Kingdom}

\author{Wenbin Li}
\affiliation{School of Engineering, Westlake University, Hangzhou 310030, China}

\author{Hua Wang}
\affiliation{Center for Quantum Matter, School of Physics, Zhejiang University, Hangzhou 310058, China.}

\author{Bartomeu Monserrat}
\email{bm418@cam.ac.uk}
\affiliation{Department of Materials Science and Metallurgy, University of Cambridge, Cambridge CB3 0FS, United Kingdom}

\date{\today}

\begin{abstract}
    Shift vectors play a central role in nonlinear optics and transport phenomena, where they are usually understood as charge-center shifts associated with transitions between quantum states. Here we show that the same geometric structure can be more fundamentally understood as the \textit{intrinsic} dipole moment of a single correlated state. Our derivation clarifies the local and global aspects of gauge invariance, the origin of the phase-gradient term, and its connection to the internal coherence structure of many-body correlations. The single-state shift character appears both as a displacement of the real-space joint probability density and as a linear electric-field modification in energy space. Applying this framework to optically induced correlations, electron-phonon-mediated processes, and excitonic electron-hole states, we recover previously proposed shift vectors and the standard expression for the shift current as special cases.
    Our results establish a common physical foundation for shift vectors as intrinsic dipolar properties of correlated electronic states.
\end{abstract}

\maketitle

\textit{Introduction.}---The shift vector has attracted broad attention because of its central role in nonlinear optical responses~\cite{belinicher1980photogalvanic,belinicher1982kinetic,von1981theory,cook2017design,sipe2000second,young2012first,qian2022role,morimoto2016topological,zhu2024anomalous} and the side-jump mechanism of anomalous Hall materials~\cite{nagaosa2010anomalous,sinitsyn2006coordinate}. Its origin can be traced to beam-shift phenomena such as the Goos-H\"anchen and Imbert-Fedorov shifts~\cite{goos1947neuer,shi2019shift}, and its relevance has since expanded to photon-drag effects~\cite{shi2021geometric}, electron--phonon coupling~\cite{belinicher1982kinetic,zhu2024anomalous,PhNLO}, exciton optical transitions~\cite{taghizadeh2018gauge,PhysRevB.101.045104,PhysRevB.109.155437,lai2024quantum,yang2025correlated,esteveparedes2025excitons}, and Landau--Zener tunneling~\cite{kitamura2020nonreciprocal}. Recent works have further emphasized its geometric significance and quantization behavior~\cite{ahn2022riemannian,wang2024geodesicnaturequantizationshift,PhysRevB.110.115108,PhysRevB.110.075159,paiva2024shift,davenport2025exciton,PhysRevB.111.L081103,PhysRevLett.133.186601}. These developments show that the shift vector is no longer a narrow concept tied to a particular optical process, but a recurring geometric quantity across a broad class of electronic responses. Across all these areas, the microscopic meaning of the shift vector is primarily formulated in a transition-based language: it is interpreted as the charge-center displacement between two states connected by a transition or scattering process.

In this Letter, we show that the shift vector can be defined more fundamentally as a property of an \textit{individual} quantum state: the generalized shift vector is the branch-fixed dipole moment of an arbitrary electronic correlated state, derived directly from its many-body wave function. In the electron-hole sector, this single-state shift appears both in the real-space joint probability density and in the linear energetic response to an applied electric field. The same framework recovers optical, excitonic, and electron-phonon shift vectors as special cases, and shows that differences between such shift vectors can probe the coherence encoded by the corresponding perturbation or interaction. This provides a common physical framework for shift vectors across different processes and many-body states.


\textit{General formulation.}---Consider an $N$-particle correlated state $S$ composed of $N_{\rm e}$ electrons and $N-N_{\rm e}$ holes,
\begin{equation}\label{eq:psi_S_generalN}
    \ket{\Psi^{S}}=\sum_{\Lambda}A^{S}_{\Lambda}\ket{\Lambda},
\end{equation}
where $\Lambda\equiv\{n_i,\sigma_i,\bm{k}_i\}_{i=1}^{N}$ labels a many-body Bloch configuration, with $\sigma_i$, $n_i$, and $\bm{k}_i$ the spin, band index, and momentum of the $i$th particle. The corresponding basis state is the direct product of $N_{\rm e}$ electron states and $N-N_{\rm e}$ hole states,
\begin{equation}
\label{eq:lambda_def}
    \ket{\Lambda}\equiv
    \prod_{i=1}^{N_{\rm e}}\ket{n_i\sigma_i,\bm{k}_i}_{\rm e}
    \prod_{i=N_{\rm e}+1}^{N}\ket{n_i\sigma_i,\bm{k}_i}_{\rm h},
\end{equation}
and the real-space wave function follows from projection onto $\ket{\{\bm{r}\}_N}=\prod_{i=1}^{N}\ket{\bm{r}_i}$ as
\begin{equation}\label{eq:psi_S_generalN_realspace}
    \begin{aligned}
        \Psi^{S}(\{\bm{r}\}_N)
        &=
        \sum_{\Lambda}
        A^{S}_{\Lambda}
        \exp\!\left[
            i\sum_{i=1}^{N}(-1)^{\zeta_i}\bm{k}_i\cdot\bm{r}_i
        \right]
        \\
        &\quad\times
        \prod_{i=1}^{N_{\rm e}}u_{n_i\sigma_i,\bm{k}_i}(\bm{r}_i)
        \prod_{i=N_{\rm e}+1}^{N}u_{n_i\sigma_i,\bm{k}_i}^{*}(\bm{r}_i),
    \end{aligned}
\end{equation}
where $\zeta_i=0$ for electrons and $\zeta_i=1$ for holes. The intrinsic dipole moment of the correlated state is
\begin{equation}\label{eq:dipole_def}
    \bm{P}^{S}
    \equiv
    \left\langle
        \Psi^{S}
        \middle|
        e\sum_{j=1}^{N}(-1)^{\zeta_j}\bm{r}_j
        \middle|
        \Psi^{S}
    \right\rangle
    \equiv
    e\,\bm{X}_{1}^{S},
\end{equation}
with $e<0$ the electron charge. As detailed in Sec.\,IA of the Supplemental Material~\cite{SI_info}, neglecting edge effects by assuming a sufficiently extended system, the direct evaluation of Eq.\,(\ref{eq:dipole_def}) gives the \textit{$N$-particle shift vector} as 
\begin{equation}\label{eq:X1_general}
    \bm{X}^{S}_{1}
    =
    \sum_{\Lambda}|A^{S}_{\Lambda}|^2
    \sum_{j=1}^{N}\nabla_{\bm{k}_j}\arg A^{S}_{\Lambda}
    -
    \left\langle
        \Psi^{S}
        \middle|
        \hat{\bm{r}}_{N}
        \middle|
        \Psi^{S}
    \right\rangle,
\end{equation}
where $\hat{\bm{r}}_N \equiv -i \sum_{j=1}^{N} \nabla_{\bm{k}_j}$ is the position operator for $N$ independent particles. Its matrix element is
\begin{equation}
    \begin{aligned}
        \left\langle \Psi^{\rm S} \middle| \hat{\bm{r}}_N \middle| \Psi^{\rm S} \right\rangle
        &=
        \sum_{\{n'_i,n_i\}}
        \sum_{\{\sigma_i,\bm{k}_i\}}
        \left( A^{\rm S}_{\{n'_i\sigma_i\bm{k}_i\}} \right)^*
        A^{\rm S}_{\{n_i\sigma_i\bm{k}_i\}} \\
        &\quad \times
        \Bigg[
            \left( \prod_{i=N_{\rm e}+1}^{N} \delta_{n'_i n_i} \right)
            \sum_{i=1}^{N_{\rm e}}
            \bm{r}_{n'_i n_i,\sigma_i,\bm{k}_i} \\
        &\qquad -
            \left( \prod_{i=1}^{N_{\rm e}} \delta_{n'_i n_i} \right)
            \sum_{i=N_{\rm e}+1}^{N}
            \bm{r}_{n_i n'_i,\sigma_i,\bm{k}_i}
        \Bigg],
    \end{aligned}
\end{equation}
with $\bm{r}_{nm,\sigma,\bm{k}} = i \langle u_{n\sigma,\bm{k}} | \nabla_{\bm{k}} u_{m\sigma,\bm{k}} \rangle$ the single-particle dipole matrix element~\cite{parker2019diagrammatic}. In the following, we absorb the spin index into the band index for notational simplicity. 


\textit{Gauge invariance.}---As a physical observable, this dipole moment $\bm{X}^{S}_{1}$ is naturally gauge invariant. This property is directly proved in Sec.\,IB of the Supplemental Material~\cite{SI_info} by considering a non-degenerate Bloch electron system with $M$ bands, in which the single-particle basis admits smooth gauge transformations of the form $\ket{n,\bm{k}} \to e^{i\chi_{n,\bm{k}}}\ket{n,\bm{k}}$, forming an Abelian subgroup $\prod_n U(1)\subset U(M)$. More generally, the geometry of a $M$-band frame is associated with the flag manifold $U(M)/U(1)^M$~\cite{ahn2022riemannian}. Extending the formulation to systems with band degeneracies or general $\bm{k}$-dependent $U(M')$ frame rotations with $1<M'<M$ requires a non-Abelian covariant treatment, which will be addressed in future work. 

Under periodic boundary conditions, $\chi_{n,\bm{k}}$ can carry a finite winding across the Brillouin zone~\cite{vanderbilt1993electric,resta1994macroscopic,PhysRevB.49.14202,vanderbilt2018berry}, which corresponds to a change of the home unit cell, or equivalently, the Wannier-center convention, or Bloch embedding~\cite{PhysRevB.89.155114,PhysRevB.110.075159,esteveparedes2025excitons}. As detailed in Sec.\,IC of the Supplemental Material~\cite{SI_info}, Eqs.~\eqref{eq:dipole_def} and \eqref{eq:X1_general} are therefore defined only up to the usual branch ambiguity. Nevertheless, for charge-neutral states with equal numbers of electrons and holes, one can choose a natural convention in which all particles are consistently referenced, for instance the center-of-mass reference, thereby fixing the branch of $\bm{X}^{S}_{1}$ (see Sec.\,I C~\cite{SI_info}).


\textit{Discussion.}---Equation~\eqref{eq:X1_general} is the central result of this work. It shows several key features that distinguish it from previously reported shift vectors and related exciton quantum-geometric formulations~\cite{belinicher1982kinetic,von1981theory,resta2024geometrical,paiva2024shift,davenport2025exciton,lai2024quantum,yang2025correlated,zhu2024anomalous}. First and foremost, it is derived directly from the electric dipole moment of a \textit{single} correlated state, rather than from a difference between two states. As a result, it depends only on the internal structure of the correlated state, encoded in the profile function $A^S_{\Lambda}$. Second, for the same reason, the formalism is not tied to any specific microscopic interaction. Instead, it applies generically to electronic correlated states that can be properly expanded in momentum space and have clear separation of electrons and holes. Third, its gauge invariance follows from the physical nature of the electric dipole itself, rather than being restored by adding compensating terms. Finally, the expression is fully multiband and does not rely on a two-band approximation. 

Benefiting from this conceptual clarity and generality, Eq.~\eqref{eq:X1_general} admits a transparent physical interpretation. The last term, $\left\langle \Psi^{\rm S} \middle| \hat{\bm{r}}_N \middle| \Psi^{\rm S} \right\rangle$, represents the dipole contribution arising from the single-particle Bloch basis. In contrast, the first term, interpreted as the averaged phase gradient of the many-body profile function $A^{S}_{\Lambda}$, is entirely induced by the internal coherence of the correlation. Such phase information is largely neglected in conventional analyses based on $|A^{S}_{\Lambda}|^2$, which primarily characterize occupation or transition probabilities. Equation~\eqref{eq:X1_general} shows that the leading observable consequence of this missing phase information is the shift vector.


\textit{Special cases.}---Several limiting cases are already instructive. The trivial case $N=0$ refers to the ``vacuum state" of the material, where any state above (below) the Fermi level is fully unoccupied (occupied). In a single-particle description with $N=N_{\rm e/h}=1$, one extra electron or hole is added as the state $|\Psi^{S}\rangle=\sum_{n\bm{k}}A^{S}_{n\bm{k}}\ket{n\bm{k}}_{\rm e/h}$, and Eq.~\eqref{eq:X1_general} reduces to the familiar gauge structure encountered in Wannierization and related changes of Bloch basis~\cite{ibanez2018ab,sjakste2015wannier}. The first nontrivial case is the two-body correlation with one electron and one hole as 
\begin{equation}\label{eq:exciton_state_conventional}
    \ket{\Psi^{S,\bm{Q}}}
    =
    \sum_{cv\bm{k}}
    A^{S,\bm{Q}}_{cv\bm{k}}
    \ket{c,\bm{k}+\bm{Q}}_{\rm e}\ket{v,\bm{k}}_{\rm h},
\end{equation}
where $\bm{Q}$ is the total momentum; $c$ and $v$ label conduction and valence bands. Equation~\eqref{eq:X1_general} then reduces to
\begin{equation}\label{eq:X1_eh_general}
    \begin{aligned}
        &\bm{X}_1^{S,\bm{Q}}
        =
        \sum_{cv\bm{k}}
        |A^{S,\bm{Q}}_{cv\bm{k}}|^2
        \nabla_{\bm{k}}\arg A^{S,\bm{Q}}_{cv\bm{k}}
        \\
        &-
        \sum_{cc',vv',\bm{k}}
        \left[
            \left(A^{S,\bm{Q}}_{c'v\bm{k}}\right)^*A^{S,\bm{Q}}_{cv\bm{k}}\bm{r}_{c'c,\bm{k}+\bm{Q}}
            -
            \left(A^{S,\bm{Q}}_{cv\bm{k}}\right)^*A^{S,\bm{Q}}_{cv'\bm{k}}
            \bm{r}_{vv',\bm{k}}
        \right], 
    \end{aligned}
\end{equation}
which directly recovers the branch-fixed exciton dipole discussed in Refs.~\cite{paiva2024shift,davenport2025exciton}, and is closely related to exciton drift velocities, maximally localized exciton Wannier constructions, and crystalline-topological shift excitons~\cite{cao2021quantum,haber2023maximally,PhysRevLett.133.176601}. Its extension to the inter-exciton-state case is also reported in the exciton-enhanced second harmonic generation~\cite{ruan2024exciton}. 

It is also worth noting that, although $\bm{X}_1^{S,\bm{Q}}$ bears a slight formal resemblance to the light-induced shift vector reported in Refs.~\cite{resta2024geometrical,yang2025correlated}, the two quantities are physically distinct. The latter characterizes the polarization difference between the excited and ground states, whereas $\bm{X}_1^{S,\bm{Q}}$ is determined solely by the internal structure of $\ket{\Psi^{S,\bm{Q}}}$. This distinction is immediate when $\ket{\Psi^{S,\bm{Q}}}$ is an exciton state formed by the Coulomb interaction. In this case, the profile function $A^{S,\bm{Q}}_{cv\bm{k}}$, which underlies the phase-gradient term, can be determined by the Bethe--Salpeter Hamiltonian $\hat{H}_{\rm BSE}$ as~\cite{albrecht1998ab,rohlfing1998electron}
\begin{equation}
    E^{S,\bm{Q}} A^{S,\bm{Q}}_{cv\bm{k}}
    =
    {}_{\rm e}\!\bra{c,\bm{k}+\bm{Q}}\,
    {}_{\rm h}\!\bra{v,\bm{k}}\,
    \hat{H}_{\rm BSE}
    \ket{\Psi^{S,\bm{Q}}},
\end{equation}
where $E^{S,\bm{Q}}$ is the exciton energy. Thus, $A^{S,\bm{Q}}_{cv\bm{k}}$ is an internal excitonic amplitude generated by
electron--hole interactions, rather than a ground-to-exciton scattering matrix
element of the type used in Refs.\,\cite{resta2024geometrical,yang2025correlated}. As discussed below, however, when the electron--hole correlation is induced externally by optical excitation, Eq.\,\eqref{eq:X1_eh_general} reduces to the light-induced shift vector of Refs.\,\cite{resta2024geometrical,yang2025correlated}. 

Furthermore, if the state is primarily spanned by a two-band subspace, Eq.~\eqref{eq:X1_general} reduces to
\begin{equation}\label{eq:X1S_twoband}
    \bm{X}_1^{S,\bm{Q}}
    =
    \sum_{\bm{k}}
    \left|A^{S,\bm{Q}}_{cv\bm{k}}\right|^2
    \bm{R}^{S,\bm{Q}}_{cv,\bm{k}},
\end{equation}
where the individual shift vector reads
\begin{equation}\label{eq:R_S_Q}
    \bm{R}^{S,\bm{Q}}_{cv,\bm{k}}
    \equiv
    \nabla_{\bm{k}}\arg A^{S,\bm{Q}}_{cv\bm{k}}
    -
    \left(
        \bm{r}_{cc,\bm{k}+\bm{Q}}
        -
        \bm{r}_{vv,\bm{k}}
    \right).
\end{equation}
If $|\Psi^{S}\rangle$ is further dominated by a single basis state at a specific $\bm{k}'$, namely $A^{S,\bm{Q}}_{cv\bm{k}}\rightarrow \delta_{\bm{k}\bm{k}'}e^{i\arg(A^{S,\bm{Q}}_{cv\bm{k}'})}$, $\bm{X}^{S,\bm{Q}}_{1}$ reduces to the individual shift vector $\bm{R}^{S,\bm{Q}}_{cv,\bm{k}'}$. In this weak-correlation limit, the phase gradient of the profile function gives the leading many-body contribution. This approximation can be valid for correlations induced by highly selective interactions, for example those with a sharp resonant energy window. Eq.\,\eqref{eq:X1S_twoband} therefore directly covers previous works on the two-band limit~\cite{xie2024theory,paiva2024shift,davenport2025exciton,zhu2024anomalous,PhysRevLett.133.176601}.


\textit{Observable consequences.}---The generalized shift vector manifests itself in two complementary ways. The most direct one is through the real-space joint probability density. To make this relation transparent, and motivated by the recent imaging of exciton wave functions~\cite{man2021experimental} and real-space exciton Wannier constructions~\cite{haber2023maximally}, we use the electron-hole state in Eq.~\eqref{eq:exciton_state_conventional} as an example, with wave function
\begin{equation}\label{eq:exciton_wavefunction_conventional}
    \begin{aligned}
        \Psi^{S}_{\bm{Q}}(\bm{r}_e,\bm{r}_h)
        &=
        \sum_{cv\bm{k}}
        A^{S,\bm{Q}}_{cv\bm{k}}
        e^{i\bm{Q}\cdot\bm{r}_e}
        e^{i\bm{k}\cdot(\bm{r}_e-\bm{r}_h)}
        \\
        &\quad\times
        u_{c,\bm{k}+\bm{Q}}(\bm{r}_e)
        u^{*}_{v,\bm{k}}(\bm{r}_h),
    \end{aligned}
\end{equation}
where $\bm{r}_e$ and $\bm{r}_h$ are the electron and hole coordinates, and $u_{c/v,\bm{k}}(\bm{r}_{\rm e/h})$ is the periodic part of single-particle wave functions. To expose the displacement encoded in $\bm{X}^{S,\bm{Q}}_{1}$, we consider the lattice-resolved exciton probability density obtained by averaging over intracell coordinates. Writing $\bm{r}_{e}=\bm{x}_{i}+\Delta\bm{r}_{e}$ and $\bm{r}_{h}=\bm{x}_{j}+\Delta\bm{r}_{h}$, with $\bm{x}_{i,j}$ the unit-cell centers and $\Delta\bm{r}_{e,h}$ the intracell coordinates, we define
\begin{equation}\label{eq:psi2_macro_localfield}
    \begin{aligned}
        \left|\Psi^S_{\bm{Q}}(\bm{x}_i-\bm{x}_j)\right|^2
        &\equiv
        \int_{\Omega}d^2\Delta\bm{r}_e
        \int_{\Omega}d^2\Delta\bm{r}_h
        \\
        &\quad\times
        \left|
            \Psi^S_{\bm{Q}}(\bm{x}_i+\Delta\bm{r}_e,\bm{x}_j+\Delta\bm{r}_h)
        \right|^2 .
    \end{aligned}
\end{equation}
As shown in Sec.\,II of the Supplemental Material~\cite{SI_info}, for two sites separated by $\bm{x}_i-\bm{x}_j=Z\bm{a}_l$, with $\bm{a}_l$ a lattice vector and $Z$ an integer, one finds
\begin{equation}
    \bm{a}_l\cdot\bm{X}^{S,\bm{Q}}_1
    \propto
    \frac{
        \left|\Psi^S_{\bm{Q}}(Z\bm{a}_l)\right|^2
        -
        \left|\Psi^S_{\bm{Q}}(-Z\bm{a}_l)\right|^2
    }{Z},
\end{equation}
which shows that $\bm{X}^{S,\bm{Q}}_1$ directly quantifies the inversion asymmetry of the electron-hole separation.

\begin{figure}[htb]
    \centering
    \includegraphics[width=0.46\textwidth]{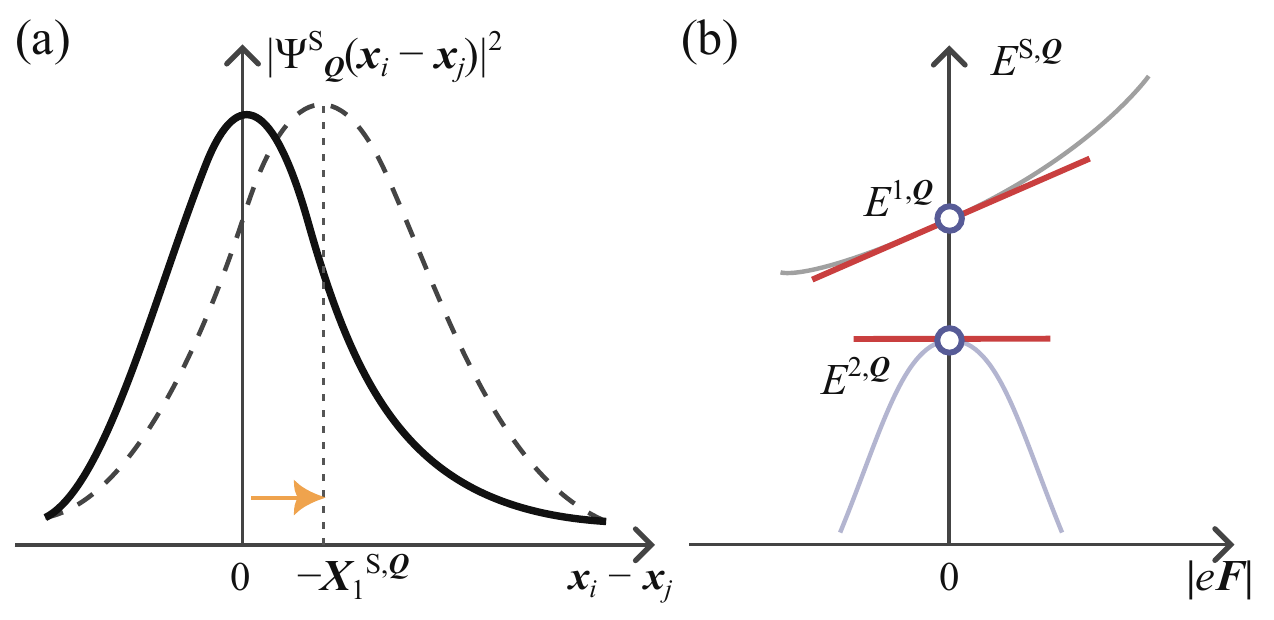}\caption{\label{fig:eh_shift_schematic}\textbf{Schematics of the physical manifestations of the exciton shift vector.} (a) Real-space shift of the (coarse-grained) probability density: the maximum peak is displaced from $\bm{x}_i-\bm{x}_j=\bm{0}$ (solid curve) to $\bm{x}_i-\bm{x}_j=-\bm{X}^{S,\bm{Q}}_1$ (dashed curve). (b) Linear Stark effect of the exciton under a homogeneous electrostatic field $\bm{F}$ in a finite system. The state $S=1$ ($S=2$) possesses a finite (vanishing) shift vector, and therefore exhibits (does not exhibit) a linear field response (red line) as the leading contribution at small $|\bm{F}|$.}
\end{figure}

To make this picture more transparent, we further consider typical two-dimensional semiconductors~\cite{qiu2013optical,chernikov2014exciton,cao2018unifying} that have dispersive and well-separated band valleys, so that low-energy excitons are localized within a small region $\Lambda$ in reciprocal space, characterized by a width $q_{\Lambda}$. In the short-range regime $|\bm{r}|\ll q_{\Lambda}^{-1}$, the coarse-grained probability density becomes
\begin{equation}\label{eq:psi2_macro_localfield_2D}
    \begin{aligned}
        \left|\Psi^S_{\bm{Q}}(\bm{x}_i-\bm{x}_j)\right|^2
        &\propto
        1
        +
        \left(
            q_{\Lambda}\left|\bm{X}_1^{S,\bm{Q}}\right|/2
        \right)^2
        \\
        &\quad-
        \frac{q_{\Lambda}^2}{8}
        \left|
            \bm{x}_i-\bm{x}_j+\bm{X}^{S,\bm{Q}}_1
        \right|^2 .
    \end{aligned}
\end{equation}
which indicates that, as schematicized in Fig.\,\ref{fig:eh_shift_schematic}(a), the maximum of the probability density is shifted from $\bm{x}_i-\bm{x}_j=\bm{0}$ to $\bm{x}_i-\bm{x}_j=-\bm{X}^{S,\bm{Q}}_1$. In this sense, the shift vector quantifies the most probable electron-hole displacement in real space, consistent with recent exciton-shift and exciton-polarization perspectives~\cite{paiva2024shift,davenport2025exciton}. This interpretation is free of gauge ambiguity and independent of any arbitrary reference position because the quantity is self-referenced and intrinsically many-body in character. It also provides a direct language for discussing inversion-asymmetric excitations~\cite{man2021experimental} and suggests a route to detection through spatial imaging of correlated states.

The second manifestation is energetic. A nonzero electric dipole $\bm{X}^{S}_1$ of the state $S$ introduces a linear-order modification to the electric enthalpy (or free energy) $\Delta \mathcal{F}^{S}$~\cite{PhysRevB.55.10337,elefield_fe,vanderbilt2018berry,cao2021quantum} in response to the external electric field $\bm{F}$, given as
\begin{equation}\label{eq:stark_shift}
    \Delta \mathcal{F}^{S}=-\bm{P}^{S}\cdot\bm{F}=-e\,\bm{X}^{S}_{1}\cdot\bm{F}.
\end{equation}
As shown schematically in Fig.\,\ref{fig:eh_shift_schematic}\,(b), in bounded systems this manifests as a linear Stark shift of the energy. This effect has been discussed and observed for excitons~\cite{scharf2016excitonic,cavalcante2018stark,leisgang2020giant}. More broadly, because $\bm{P}^{S}=e\bm{X}^{S}_{1}$ is the intrinsic dipole moment of the correlated state, it also governs physical effects that couple directly to an electric dipole, such as electrostatic confinement at heterojunctions~\cite{thureja2022electrically}. Recent work further shows that this dipolar quantity remains robust when the correlation is localized in real space~\cite{yang2025correlated}.



\textit{Interaction fingerprints.}---Because the phase-gradient contribution to $\bm{X}^{S}_{1}$ is inherited from the profile function, the shift vector can also be used to probe the interaction that creates or transforms a correlated state. As an example, consider a monochromatic optical electric field $\bm{E}_0\cos(\Omega t)e^{-\gamma|t|}$ with central frequency $\Omega$ and decay rate $\gamma$. In the interaction picture, the coupling operator is
\begin{equation}
    \hat{V}_I(t)
    =
    e\,\hat{\bm{r}}(t)\cdot\bm{e}_p
    |\bm{E}_0|
    \cos(\Omega t)e^{-\gamma|t|},
\end{equation}
where $\hat{\bm{r}}$ is the position operator and $\bm{e}_p$ is the polarization direction. As detailed in Sec.\,III of the Supplemental Material~\cite{SI_info}, in the low-temperature limit and keeping only the first-order perturbation of $\hat{V}_I(t)$, the resonant contribution evolves the ground state $|0\rangle$ to
\begin{equation}
    \begin{aligned}
        \ket{\Psi^{\Omega,\bm{E}_0}(t)}
        &\approx
        \ket{0}
        +
        \frac{i e|\bm{E}_0|}{\hbar}
        \left(
            \frac{2-e^{-\gamma t}}{2\gamma}
        \right)
        \\
        &\quad\times
        \sum_{cv\bm{k}\in C_{\Omega}}
        \bm{e}_p\cdot\bm{r}_{cv,\bm{k}}
        \ket{c,\bm{k}}_{\rm e}\ket{v,\bm{k}}_{\rm h}.
    \end{aligned}
\end{equation}
where the set $C_{\Omega}$ includes bands satisfying $\omega_{cv,\bm{k}}=|\Omega|$, with $\omega_{cv,\bm{k}}\equiv(\epsilon_{c,\bm{k}}-\epsilon_{v,\bm{k}})/\hbar$ the interband transition frequency. This is the electron-hole correlated state generated by the optical excitation. Comparing with Eq.~\eqref{eq:psi_S_generalN}, the profile function is directly proportional to $\bm{e}_p\cdot\bm{r}_{cv,\bm{k}}$. Therefore, neglecting band degeneracies, the intrinsic dipole of this state follows from Eq.~\eqref{eq:X1S_twoband} as
\begin{equation}
    \begin{aligned}
        e\bm{X}^{\Omega,\bm{E}_0}_1
        &=
        \frac{e^3|\bm{E}_0|^2}{\hbar^2}
        \left(
            \frac{2-e^{-\gamma t}}{2\gamma}
        \right)^2
        \\
        &\quad\times
        \sum_{cv\bm{k}\in C_{\Omega}}
        \left|\bm{e}_p\cdot\bm{r}_{cv,\bm{k}}\right|^2
        \bm{R}^{\bm{e}_p}_{cv,\bm{k}},
    \end{aligned}
\end{equation}
where the optical shift vector appears in the conventional form
\begin{equation}\label{eq:R_optic}
    \bm{R}^{\bm{e}_p}_{cv,\bm{k}}
    =
    \bm{r}_{cc,\bm{k}}
    -
    \bm{r}_{vv,\bm{k}}
    -
    \nabla_{\bm{k}}
    \arg(\bm{e}_p\cdot\bm{r}_{cv,\bm{k}}).
\end{equation}
In the steady-illumination limit $\gamma\rightarrow0$, the leading temporal variation of this dipole yields the direct current as
\begin{equation}
    \begin{aligned}
        \bm{J}_{\rm dc}
        &=
        \lim_{\gamma \rightarrow 0}\frac{d}{dt}(e\bm{X}^{\Omega,\bm{E}_0}_1)
        \\
        &\rightarrow -\frac{e^3|\bm{E}_0|^2}{2\hbar^2}
        \sum_{cv\bm{k}}
        \left|\bm{e}_p\cdot\bm{r}_{cv,\bm{k}}\right|^2
        \bm{R}^{\bm{e}_p}_{cv,\bm{k}}
        \delta(\omega_{cv,\bm{k}}-|\Omega|),
    \end{aligned}
\end{equation}
This provides a transparent interpretation of the phase-gradient term $\nabla_{\bm{k}}\arg(\bm{e}_p\cdot\bm{r}_{cv,\bm{k}})$: it is the coherence contribution carried by the optically generated electron-hole correlation. Note that this is not the interacting exciton shift current~\cite{taghizadeh2018gauge,PhysRevB.101.045104,PhysRevB.109.155437,nakamura2024strongly,chan2021giant,esteveparedes2025excitons} since the Coulomb interaction between electron and hole is not considered. This recovery of the standard shift-current expression~\cite{belinicher1980photogalvanic,belinicher1982kinetic,von1981theory,sipe2000second,young2012first,cook2017design} should be read as a benchmark of the general construction, not as an identification of the steady shift current with an equilibrium single-state observable. Rather, the optical field prepares a driven electron-hole correlation whose dipolar displacement factor is precisely the conventional optical shift vector. 

The same logic extends to a general monochromatic potential of the form
\begin{equation}
    \hat{V}_I(t)
    =
    V_{n\bm{k}_1,m\bm{k}_2}
    \hat{c}^{\dagger}_{n\bm{k}_1}\hat{c}_{m\bm{k}_2}
    e^{i(\Omega-\omega_{n\bm{k}_1,m\bm{k}_2})t},
\end{equation}
which generates electron-hole correlations with profile function proportional to $V_{c\bm{k}_1,v\bm{k}_2}$. This suggests a second scheme for probing interactions: compare two correlated states induced by different interactions. If two interactions $\hat{V}_1$ and $\hat{V}_2$ are both highly selective to individual electron-hole pairs, their weak-hybridization limit difference reads
\begin{equation}
    \bm{R}^{V_1,\bm{Q}}_{cv,\bm{k}}
    -
    \bm{R}^{V_2,\bm{Q}}_{cv,\bm{k}}
    =
    \nabla_{\bm{k}}
    \arg\!\left(
        \frac{V^1_{c(\bm{k}+\bm{Q}),v\bm{k}}}
             {V^2_{c(\bm{k}+\bm{Q}),v\bm{k}}}
    \right),
\end{equation}
leaving only the phase difference and therefore offering an accessible way to detect the coherence of one interaction relative to another.

As an example, electron-phonon coupling to a phonon mode $\lambda$ with momentum $\bm{Q}$ produces the matrix element $\langle cv,\bm{k},\bm{Q}|\hat{g}^{\lambda,\bm{Q}}|0\rangle=g^{\lambda,\bm{Q}}_{c(\bm{k}+\bm{Q}),v\bm{k}}$, and therefore the shift vector
\begin{equation}
    \bm{R}^{\lambda,\bm{Q}}_{cv,\bm{k}}
    =
    \bm{r}_{cc,\bm{k}+\bm{Q}}
    -
    \bm{r}_{vv,\bm{k}}
    -
    \nabla_{\bm{k}}
    \arg g^{\lambda,\bm{Q}}_{c(\bm{k}+\bm{Q}),v\bm{k}}.
\end{equation}
Its difference from the optical shift vector was identified in our previous work as the driving quantity of the phonon-mediated shift current~\cite{PhNLO}. As another example, at the optical limit $\bm{Q} = 0^+\bm{e}_{\alpha}$ with $\bm{e}_{\alpha}$ the unit vector along $\alpha$ direction, the exciton shift vector defined in Eq.~\eqref{eq:R_S_Q} differs from the optical shift vector as
\begin{equation}\label{eq:R_ex_optical}
    \begin{aligned}
        \bm{R}^{S,0^+\bm{e}_{\alpha}}_{cv,\bm{k}} - \bm{R}^{\alpha}_{cv,\bm{k}}
        &= - \nabla_{\bm{k}} \arg\left(A^{S,\bm{0}}_{cv,\bm{k}}r^{\alpha}_{vc}\right) 
    \end{aligned}
\end{equation}
which encodes the internal phase structure of the exciton dipole $\bm{M}^{S,\bm{0}} = \sum_{\bm{k}} A^{S,\bm{0}}_{cv,\bm{k}}\bm{r}_{vc}$. 
This structure quantifies how the Coulomb interaction coherently hybridizes free electron--hole pairs into collective excitations: according to Eq.\,\eqref{eq:X1S_twoband}, in the trivial limit where the right-hand side of Eq.\,\eqref{eq:R_ex_optical} vanishes, the total exciton shift vector $\bm{X}^{S,0^+\bm{e}_{\alpha}}_1$ becomes simply the classical average of the optical shift vectors of the constituent free electron--hole pairs. In contrast, for nontrivial phase structure, the exciton shift vector can no longer be factorized into individual contributions from free electron-hole pairs. 

\begin{figure}[htb]
    \centering
    \includegraphics[width=0.46\textwidth]{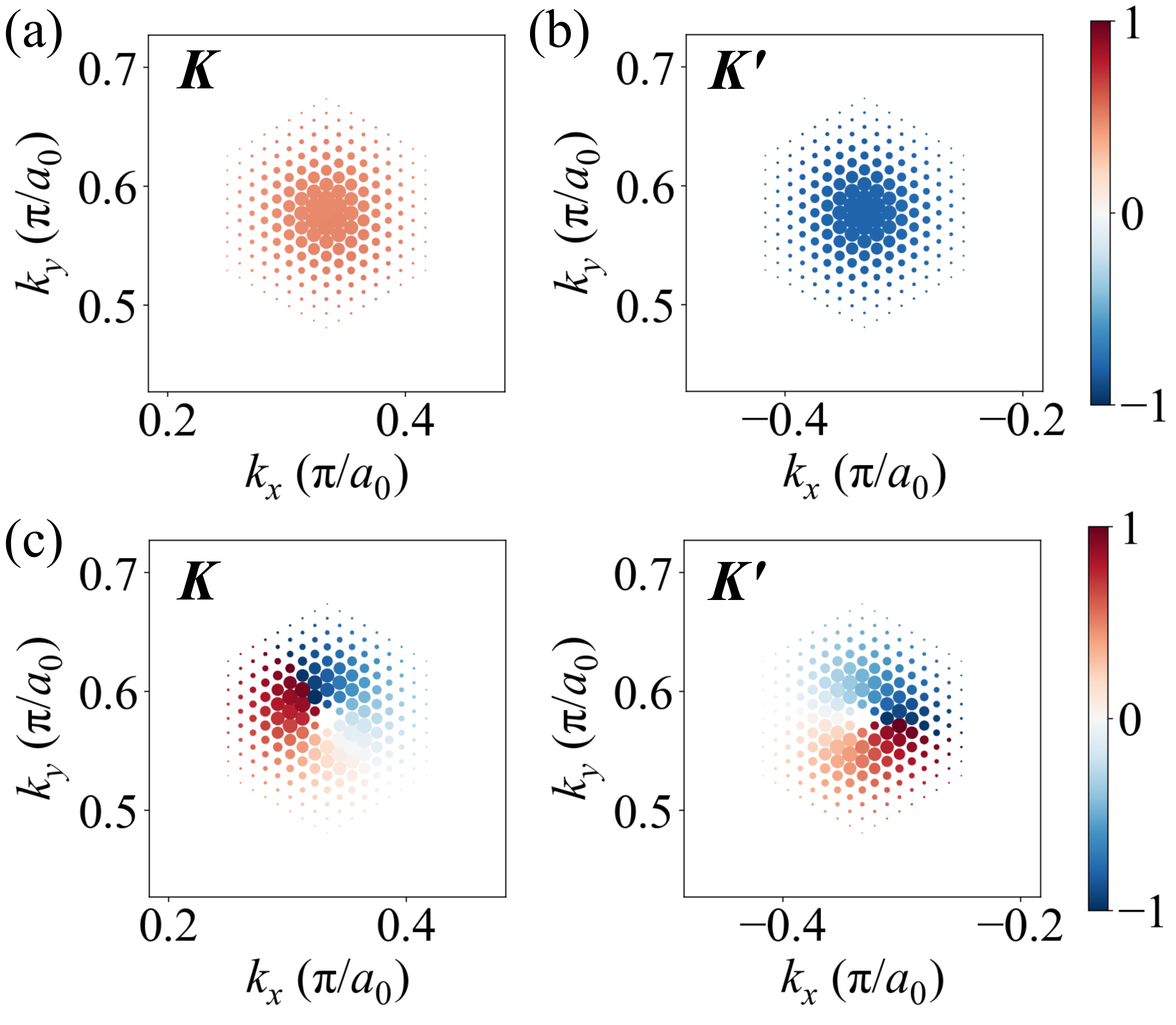}
    \caption{\label{fig:MoS2_7_8_11} \textbf{First principles description of excitons in monolayer MoS$_2$.} The internal phase of $x$-direction exciton dipole $\arg(A^{S,0}_{cv\bm{k}}\, r^{b_1}_{cv,\bm{k}})$ of (a) 7th (b) 8th and (c) 11th lowest-energy exciton states (unit: $\pi$; $b_1$ aligns with one of the lattice vectors $(\sqrt{3},-1)$). The size of dots is scaled by $|A^{S,0}_{cv\bm{k}}|^2$. The profile function of low-energy excitons are strongly localized near $\bm{K}$ and (or) $\bm{K}'$ valleys in $\bm{k}$ space, and primarily composed of the lowest conduction band $c$ and the highest valence band $v$ summed over both spins. }
\end{figure}

As an illustrative example, first-principles results for monolayer MoS$_2$ are shown in Fig.\,\ref{fig:MoS2_7_8_11}. We consider the 7th, 8th, and 11th lowest-energy excitons as representative cases, which are two bright 1$s$- and one 2$p$-like excitons for in-plane polarized light~\cite{qiu2016screening}. The former two display a trivial phase structure, where $\arg\!\left(A^{S,\bm{0}}_{cv,\bm{k}} r^{b_1}_{vc}\right)$ remains nearly constant. In contrast, the latter exhibits a finite phase gradient $\nabla_{\bm{k}} \arg\!\left(A^{S,\bm{0}}_{cv,\bm{k}} r^{b_1}_{vc}\right) \neq 0$. Nontrivial exciton phase textures and the associated exciton geometric structures have been reported in previous works~\cite{cao2018unifying,Signatures166802,zhou2015berry,xie2024theory,paiva2024shift,davenport2025exciton,lai2024quantum}. 
Here, however, Eq.~\eqref{eq:R_ex_optical} highlights them for the first time as a manifestation of correlation coherence. 
This relation also provides a practical first-principles route for evaluating exciton shift vectors fully through gauge-invariant ingredients. 


\textit{Other applications.}---The present framework is not restricted to excitons. For electron-electron or hole-hole correlations, the same construction yields the intrinsic relative displacement of the correlated pair. For instance, as detailed in Sec.\,IV of the Supplemental Material~\cite{SI_info}, the Cooper-pair state can be viewed as an ``effective electron-hole'' superposition in time-reversal symmetric systems, and the shift vector $\bm{X}^{S}_{1}$ represents the average separation between two electrons. More generally, charged and neutral higher-body excitonic eigenstates can be treated on the same footing by applying Eq.~\eqref{eq:X1_general} to their profile functions. Representative examples include trions~\cite{deilmann2016threeparticle}, biexcitons and trion-exciton complexes~\cite{li2018revealing}, and intercell moir\'e exciton complexes in correlated moir\'e charge backgrounds~\cite{wang2023intercell}. For correlated states in quantum Hall systems, where explicit electron and hole degrees of freedom are not readily identifiable, the intrinsic dipole can instead be defined through a density-matrix construction~\cite{PhysRevB.111.035158}.


\textit{Conclusion.}---We have shown that the generalized shift vector gives the branch-fixed intrinsic dipole of an electronic correlated state, and controls both the real-space probability shift and the linear electric-field response. This result clarifies the physical meaning of the many-body formula, explains the phase-gradient term as the indispensable contribution of coherent correlation, and unifies phenomena that are often discussed separately, ranging from excitons to photocurrent. The formulation is general, gauge invariant, and directly transferable to realistic many-body calculations. It therefore provides a compact route for analyzing internal coherence in correlated electronic excitations.


\textit{Acknowledgements.}---The authors thank Prof. Raffaele Resta from IOM-CNR, Prof. Robert-Jan Slager from the University of Manchester, Prof. Justin Song from NTU Singapore, Prof. Herbert Fertig from Indiana University, Prof. Diana Qiu and Mr Xian Xu from Yale University, Mr Wojciech J. Jankowski and Dr Ruoshi Jiang from the University of Cambridge, and Prof. Likun Shi from Zhejiang University for insightful discussions. J.H. acknowledges the support of the China Scholarship Council. S.K. and B.M. are supported by an EPSRC grant [EP/V062654/1]. J.J.P.T. and B.M. are supported by an EPSRC Programme grant [EP/W017091/1]. W.L. acknowledges support from the National Natural Science Foundation of China (NSFC) under Project No. 62374136. H.W. acknowledges the support from the NSFC under Grants Nos. 12522411, 12474240, and 12304049. B.M. is also supported by a UKRI Future Leaders Fellowship [UKRI2083].

\let\savedaddcontentsline\addcontentsline
\renewcommand{\addcontentsline}[3]{}
\bibliography{references}
\let\addcontentsline\savedaddcontentsline

\clearpage
\onecolumngrid
\setcounter{section}{0}
\setcounter{equation}{0}
\setcounter{figure}{0}
\setcounter{table}{0}
\renewcommand{\theequation}{S\arabic{equation}}
\renewcommand{\thefigure}{S\arabic{figure}}
\renewcommand{\thetable}{S\arabic{table}}
\renewcommand{\theHequation}{S\arabic{equation}}
\renewcommand{\theHfigure}{S\arabic{figure}}
\renewcommand{\theHtable}{S\arabic{table}}

\begin{center}
{\large\bfseries Supplemental Material for\\[0.5em]
``Generalized Shift Vector as the Intrinsic Dipole of Many-Body Correlated Electronic States''}
\end{center}

\tableofcontents

\section{Derivation of the \texorpdfstring{$N$}{N}-Particle Shift Vector}\label{sec:SI_general}

\subsection{Many-body basis and direct evaluation}\label{sec:SI_general_derivation}

A correlated electronic state $S$ with $N_{\rm e}$ electrons and $N-N_{\rm e}$ holes can be expanded on the Bloch-product basis as
\begin{equation}\label{eq:SI_psi_S_generalN}
    \ket{\Psi^{S}}
    =
    \sum_{\Lambda}
    A^{S}_{\Lambda}\ket{\Lambda},
\end{equation}
where
\begin{equation}\label{eq:SI_lambda_def}
    \Lambda
    \equiv
    \{n_i,\sigma_i,\bm{k}_i\}_{i=1}^{N},
    \qquad
    \ket{\Lambda}
    \equiv
    \prod_{i=1}^{N_{\rm e}}\ket{n_i\sigma_i,\bm{k}_i}_{\rm e}
    \prod_{i=N_{\rm e}+1}^{N}\ket{n_i\sigma_i,\bm{k}_i}_{\rm h}.
\end{equation}
Here $n_i$, $\sigma_i$, and $\bm{k}_i$ label the band, spin, and momentum of the $i$th particle. The wave function of the single-particle basis states is given as
\begin{equation}
    \langle \bm{r}| n\sigma,\bm{k}\rangle_{\rm e}
    =
    e^{i\bm{k}\cdot\bm{r}}u_{n\sigma,\bm{k}}(\bm{r}),
    \qquad
    \langle \bm{r}| n\sigma,\bm{k}\rangle_{\rm h}
    =
    e^{-i\bm{k}\cdot\bm{r}}u^*_{n\sigma,\bm{k}}(\bm{r}),
\end{equation}
with $u_{n\sigma,\bm{k}}(\bm{r})$ the cell-periodic part of the Bloch wave function. Projection onto the $N$-particle position basis $\ket{\{\bm{r}\}_N}\equiv\prod_{i=1}^{N}\ket{\bm{r}_i}$ gives
\begin{equation}\label{eq:SI_psi_S_generalN_realspace}
    \begin{aligned}
        \Psi^{S}(\{\bm{r}\}_N)
        &\equiv
        \langle \{\bm{r}\}_N \mid \Psi^{S}\rangle \\
        &=
        \sum_{\Lambda}
        A^{S}_{\Lambda}
        \exp\!\left[
            i\sum_{i=1}^{N}(-1)^{\zeta_i}\bm{k}_i\cdot\bm{r}_i
        \right]
        \prod_{i=1}^{N_{\rm e}}u_{n_i\sigma_i,\bm{k}_i}(\bm{r}_i)
        \prod_{i=N_{\rm e}+1}^{N}u^*_{n_i\sigma_i,\bm{k}_i}(\bm{r}_i),
    \end{aligned}
\end{equation}
where $\zeta_i=0$ for electrons and $\zeta_i=1$ for holes. Its absolute value squared $|\Psi^{S}(\{\bm{r}\}_N)|^2$ gives the joint probability density of finding particles at specific positions $\{\bm{r}_i|i=1,\ldots,N\}$ in real space. In a $d$-dimensional system, it is normalized over the whole $\mathbb{R}^{dN}$ space as 
\begin{equation}
    \langle \Psi^S | \Psi^S \rangle = \prod_{i=1}^N \int d^d\bm{r}_i  |\Psi^{S}(\{\bm{r}\}_N)|^2 = 1. 
\end{equation}
In general it cannot be factorized into separate electron and hole probability densities, reflecting the presence of electron-hole correlations. 

The intrinsic dipole is obtained from the operator $e\sum_{j=1}^{N}(-1)^{\zeta_j}\bm{r}_j$. It is convenient to first act with the dipole operator on the wave function: 
\begin{equation}\label{eq:SI_half_dipole_density}
    \begin{aligned}
        &\left[ \sum_{j=1}^{N}(-1)^{\zeta_j}\bm{r}_j \right] \Psi^{S}(\{\bm{r}\}_N) \\
        &=
        \sum_{\Lambda}
        \sum_{j=1}^{N}(-1)^{\zeta_j}\bm{r}_j
        e^{i\sum_{i=1}^{N}(-1)^{\zeta_i}\bm{k}_i\cdot\bm{r}_i}
        A^{S}_{\Lambda}
        \prod_{i=1}^{N_{\rm e}}u_{n_i\sigma_i,\bm{k}_i}(\bm{r}_i)
        \prod_{i=N_{\rm e}+1}^{N}u^*_{n_i\sigma_i,\bm{k}_i}(\bm{r}_i) \\
        &=
        -i
        \sum_{\Lambda}
        \left[
            \sum_{j=1}^{N}\nabla_{\bm{k}_j}
            e^{i\sum_{i=1}^{N}(-1)^{\zeta_i}\bm{k}_i\cdot\bm{r}_i}
        \right]
        A^{S}_{\Lambda}
        \prod_{i=1}^{N_{\rm e}}u_{n_i\sigma_i,\bm{k}_i}(\bm{r}_i)
        \prod_{i=N_{\rm e}+1}^{N}u^*_{n_i\sigma_i,\bm{k}_i}(\bm{r}_i) \\
        &=
        i
        \sum_{\Lambda}
        e^{i\sum_{i=1}^{N}(-1)^{\zeta_i}\bm{k}_i\cdot\bm{r}_i}
        \sum_{j=1}^{N}\nabla_{\bm{k}_j}
        \left[
            A^{S}_{\Lambda}
            \prod_{i=1}^{N_{\rm e}}u_{n_i\sigma_i,\bm{k}_i}(\bm{r}_i)
            \prod_{i=N_{\rm e}+1}^{N}u^*_{n_i\sigma_i,\bm{k}_i}(\bm{r}_i)
        \right].
    \end{aligned}
\end{equation}
In the last equality we have integrated by parts in momentum space, which is justified by the periodic boundary conditions in $\bm{k}$-space. Multiplying by ${\Psi^S}^*(\{\bm{r}\}_N)$ yields the dipole density 
\begin{equation}\label{eq:SI_dipole_density_full}
    \begin{aligned}
        &{\Psi^S}^*(\{\bm{r}\}_N)
        \left[ \sum_{j=1}^{N}(-1)^{\zeta_j}\bm{r}_j \right] \Psi^S(\{\bm{r}\}_N) \\
        &=
        i
        \sum_{\Lambda',\Lambda}
        \exp\!\left[
            i\sum_{i=1}^{N}(-1)^{\zeta_i}
            (\bm{k}_i-\bm{k}'_i)\cdot\bm{r}_i
        \right]
        \left[
            A^{S}_{\Lambda'}
            \prod_{i=1}^{N_{\rm e}}u_{n'_i\sigma'_i,\bm{k}'_i}(\bm{r}_i)
            \prod_{i=N_{\rm e}+1}^{N}u^*_{n'_i\sigma'_i,\bm{k}'_i}(\bm{r}_i)
        \right]^* \\
        &\quad\times
        \sum_{j=1}^{N}\nabla_{\bm{k}_j}
        \left[
            A^{S}_{\Lambda}
            \prod_{i=1}^{N_{\rm e}}u_{n_i\sigma_i,\bm{k}_i}(\bm{r}_i)
            \prod_{i=N_{\rm e}+1}^{N}u^*_{n_i\sigma_i,\bm{k}_i}(\bm{r}_i)
        \right].
    \end{aligned}
\end{equation}
We then assume no edge effects by considering a sufficiently extended system, or equivalently, a sufficiently localized $|\Psi^{\rm S}\rangle$ in real space compared with the system size. Therefore, the oscillatory factor in Eq.\,\eqref{eq:SI_dipole_density_full} vanishes upon integration over all unit cells unless $\bm{k}'_i=\bm{k}_i$ for every particle. Keeping only this homogeneous contribution, we obtain 
\begin{equation}\label{eq:SI_dipole_density_homogeneous}
    \begin{aligned}
        &{\Psi^S}^*(\{\bm{r}\}_N)
        \left[ \sum_{j=1}^{N}(-1)^{\zeta_j}\bm{r}_j \right]\Psi^S(\{\bm{r}\}_N) 
        \\
        &\rightarrow
        i
        \sum_{\{n'_i,\sigma'_i\}}
        \sum_{\Lambda}
        \left(A^{S}_{\{n'_i,\sigma'_i,\bm{k}_i\}}\right)^*
        \sum_{j=1}^{N}\nabla_{\bm{k}_j}
        A^{S}_{\Lambda}
        \prod_{i=1}^{N_{\rm e}}
        \rho_{n'_in_i,\sigma'_i\sigma_i,\bm{k}_i}(\bm{r}_i)
        \prod_{i=N_{\rm e}+1}^{N}
        \rho^*_{n'_in_i,\sigma'_i\sigma_i,\bm{k}_i}(\bm{r}_i) 
        \\
        &+
        i
        \sum_{\{n'_i,\sigma'_i\}}
        \sum_{\Lambda}
        \left(A^{S}_{\{n'_i,\sigma'_i,\bm{k}_i\}}\right)^*
        A^{S}_{\Lambda}
        \sum_{j=1}^{N_{\rm e}}
        u^*_{n'_j\sigma'_j,\bm{k}_j}(\bm{r}_j)
        \nabla_{\bm{k}_j}u_{n_j\sigma_j,\bm{k}_j}(\bm{r}_j)
        \\
        &{\times}
        \prod_{i=1, i\neq j}^{N_{\rm e}}
        \rho_{n'_in_i,\sigma'_i\sigma_i,\bm{k}_i}(\bm{r}_i)
        \prod_{i=N_{\rm e}+1}^{N}
        \rho^*_{n'_in_i,\sigma'_i\sigma_i,\bm{k}_i}(\bm{r}_i) 
        \\
        &+
        i
        \sum_{\{n'_i,\sigma'_i\}}
        \sum_{\Lambda}
        \left(A^{S}_{\{n'_i,\sigma'_i,\bm{k}_i\}}\right)^*
        A^{S}_{\Lambda}
        \sum_{j=N_{\rm e}+1}^{N}
        u_{n'_j\sigma'_j,\bm{k}_j}(\bm{r}_j)
        \nabla_{\bm{k}_j}u^*_{n_j\sigma_j,\bm{k}_j}(\bm{r}_j)
        \\
        &{\times}
        \prod_{i=1}^{N_{\rm e}}
        \rho_{n'_in_i,\sigma'_i\sigma_i,\bm{k}_i}(\bm{r}_i)
        \prod_{\substack{i=N_{\rm e}+1 \\ i\neq j}}^{N}
        \rho^*_{n'_in_i,\sigma'_i\sigma_i,\bm{k}_i}(\bm{r}_i),
    \end{aligned}
\end{equation}
where
\begin{equation}
    \rho_{n'n,\sigma'\sigma,\bm{k}}(\bm{r})
    \equiv
    u^*_{n'\sigma',\bm{k}}(\bm{r})u_{n\sigma,\bm{k}}(\bm{r}).
\end{equation}
The Bloch-basis completeness relation gives
\begin{equation}\label{eq:SI_grad_cell_wavefunction}
    \begin{aligned}
        u^*_{n'\sigma,\bm{k}}(\bm{r})\nabla_{\bm{k}}u_{n\sigma,\bm{k}}(\bm{r})
        &=
        -i\sum_{m}
        \rho_{n'm,\sigma\sigma,\bm{k}}(\bm{r})
        \bm{r}_{mn,\sigma,\bm{k}}, \\
        u_{n'\sigma,\bm{k}}(\bm{r})\nabla_{\bm{k}}u^*_{n\sigma,\bm{k}}(\bm{r})
        &=
        i\sum_{m}
        \rho^*_{n'm,\sigma\sigma,\bm{k}}(\bm{r})
        \bm{r}_{nm,\sigma,\bm{k}},
    \end{aligned}
\end{equation}
with the single-particle dipole matrix element
\begin{equation}
    \bm{r}_{nm,\sigma,\bm{k}}
    \equiv
    i\langle u_{n\sigma,\bm{k}} \mid \nabla_{\bm{k}}u_{m\sigma,\bm{k}}\rangle.
\end{equation}


Equation\,\eqref{eq:SI_dipole_density_homogeneous} is periodic over the unit cell. Accordingly, the expectation value $\bm{P}^S$ can be obtained by integrating this quantity over the unit cell with respect to each particle coordinate. Using $\int_{\Omega}d^d\bm{r}\,\rho_{n'n,\sigma'\sigma,\bm{k}}(\bm{r})=\delta_{n'n}\delta_{\sigma'\sigma}$, we find
\begin{equation}\label{eq:SI_dipole_general}
    \begin{aligned}
        \bm{P}^S
        &\equiv
        \left\langle
            \Psi^{S}
            \middle|
            e\sum_{j=1}^{N}(-1)^{\zeta_j}\bm{r}_j
            \middle|
            \Psi^{S}
        \right\rangle
        \\
        &=
        e\prod_{j=1}^{N}\int_{\Omega}d^d\bm{r}_j\,
        {\Psi^S}^*(\{\bm{r}\}_N)
        \left[\sum_{j=1}^{N}(-1)^{\zeta_j}\bm{r}_j\right]
        \Psi^S(\{\bm{r}\}_N) \\
        &=
        e
        \sum_{\Lambda}
        |A^{S}_{\Lambda}|^2
        \sum_{j=1}^{N}\nabla_{\bm{k}_j}\arg A^{S}_{\Lambda}
        -
        e
        \left\langle
            \Psi^{S}
            \middle|
            \hat{\bm{r}}_N
            \middle|
            \Psi^{S}
        \right\rangle
        \equiv
        e\,\bm{X}^{S}_{1},
    \end{aligned}
\end{equation}
where
\begin{equation}
    \hat{\bm{r}}_N
    \equiv
    -i\sum_{j=1}^{N}\nabla_{\bm{k}_j}
\end{equation}
is the independent-particle position operator in the Bloch basis. Its matrix element can be written as
\begin{equation}\label{eq:SI_rN_matrix_element}
    \begin{aligned}
        \left\langle
            \Psi^{S}
            \middle|
            \hat{\bm{r}}_N
            \middle|
            \Psi^{S}
        \right\rangle
        &=
        \sum_{\{n'_i,n_i\}}
        \sum_{\{\sigma_i,\bm{k}_i\}}
        \left(A^{S}_{\{n'_i\sigma_i\bm{k}_i\}}\right)^*
        A^{S}_{\{n_i\sigma_i\bm{k}_i\}} \\
        &\quad\times
        \Bigg[
            \left(
                \prod_{i=N_{\rm e}+1}^{N}\delta_{n'_in_i}
            \right)
            \sum_{i=1}^{N_{\rm e}}
            \bm{r}_{n'_in_i,\sigma_i,\bm{k}_i} \\
        &\qquad-
            \left(
                \prod_{i=1}^{N_{\rm e}}\delta_{n'_in_i}
            \right)
            \sum_{i=N_{\rm e}+1}^{N}
            \bm{r}_{n_in'_i,\sigma_i,\bm{k}_i}
        \Bigg].
    \end{aligned}
\end{equation}
Equation~\eqref{eq:SI_dipole_general} uses the identity
\begin{equation}\label{eq:SI_real_part_identity}
    \sum_{\Lambda}
    \mathrm{Re}\!\left[
        \left(A^{S}_{\Lambda}\right)^*
        \sum_{j=1}^{N}\nabla_{\bm{k}_j}A^{S}_{\Lambda}
    \right]
    =
    \frac{1}{2}
    \sum_{\Lambda}
    \sum_{j=1}^{N}\nabla_{\bm{k}_j}|A^{S}_{\Lambda}|^2
    =
    0,
\end{equation}
which follows from normalization together with periodic boundary conditions in momentum space.


\subsection{Gauge invariance}\label{sec:SI_gauge}

Under a smooth $U(1)$ gauge transformation, each Bloch state transforms as
\begin{equation}
    \ket{n,\sigma,\bm{k}}
    \rightarrow
    e^{i\chi_{n\sigma,\bm{k}}}\ket{n,\sigma,\bm{k}}, 
\end{equation}
thereby the real-space wave function in Eq.\,\eqref{eq:SI_psi_S_generalN_realspace} changes as
\begin{equation}
    \Psi^{S}(\{\bm{r}\}_N)
    \rightarrow
    e^{i\chi_{S}}
    \Psi^{S}(\{\bm{r}\}_N)
\end{equation}
provided the profile function transforms covariantly,
\begin{equation}\label{eq:SI_gauge_trans_A}
    A^{S}_{\Lambda}
    \rightarrow
    A^{S}_{\Lambda}
    \exp\!\left[
        -i\sum_{i=1}^{N}(-1)^{\zeta_i}\chi_{n_i\sigma_i,\bm{k}_i}
    \right]
    e^{i\chi_S},
\end{equation}
with $\chi_S$ an overall phase of the many-body state.

Equation~\eqref{eq:SI_gauge_trans_A} implies
\begin{equation}
    \begin{aligned}
        &\sum_{\Lambda}
        |A^{S}_{\Lambda}|^2
        \sum_{j=1}^{N}\nabla_{\bm{k}_j}\arg A^{S}_{\Lambda} \\
        &\rightarrow
        \sum_{\Lambda}
        |A^{S}_{\Lambda}|^2
        \sum_{j=1}^{N}\nabla_{\bm{k}_j}\arg A^{S}_{\Lambda}
        -
        \sum_{\Lambda}
        |A^{S}_{\Lambda}|^2
        \sum_{j=1}^{N}(-1)^{\zeta_j}\nabla_{\bm{k}_j}\chi_{n_j\sigma_j,\bm{k}_j}.
    \end{aligned}
\end{equation}
The single-particle dipole matrix element transforms as
\begin{equation}
    \bm{r}_{nm,\sigma,\bm{k}}
    \rightarrow
    \bm{r}_{nm,\sigma,\bm{k}}
    -
    \delta_{nm}\nabla_{\bm{k}}\chi_{n\sigma,\bm{k}},
\end{equation}
and therefore
\begin{equation}
    \begin{aligned}
        \left\langle
            \Psi^{S}
            \middle|
            \hat{\bm{r}}_N
            \middle|
            \Psi^{S}
        \right\rangle
        &\rightarrow
        \left\langle
            \Psi^{S}
            \middle|
            \hat{\bm{r}}_N
            \middle|
            \Psi^{S}
        \right\rangle \\
        &\quad-
        \sum_{\Lambda}
        |A^{S}_{\Lambda}|^2
        \sum_{j=1}^{N}(-1)^{\zeta_j}\nabla_{\bm{k}_j}\chi_{n_j\sigma_j,\bm{k}_j}.
    \end{aligned}
\end{equation}
The two gauge-dependent terms cancel exactly, so $\bm{X}^{S}_{1}$ is invariant. 



\subsection{Branch ambiguity}\label{sec:SI_branch}

The gauge-invariance proof above is local in character: it assumes a fixed choice of real-space unit cell, or equivalently a fixed polarization branch. As discussed in the modern theory of polarization~\cite{vanderbilt1993electric,resta1994macroscopic,PhysRevB.49.14202,vanderbilt2018berry}, this branch ambiguity can be seen directly in the real-space definition. For simplicity, consider one particle in one spatial dimension and choose the unit cell $\Omega=[0,a)$. Equation~\eqref{eq:SI_dipole_general} then reduces to
\begin{equation}
    P^S(\Omega)
    =
    e\int_{\Omega} dx\, x\, |\Psi^S(x)|^2 .
\end{equation}
If instead we represent the same periodic state using the shifted unit cell $\Omega_R=[R,R+a)$ with $R=ma$ a lattice vector, then periodicity of $|\Psi^S(x)|^2$ gives
\begin{equation}
    \begin{aligned}
        P^S(\Omega_R)
        =
        e\int_{\Omega_R} dx\, x\, |\Psi^S(x)|^2 =
        e\int_{\Omega} dx\, (x+R)\, |\Psi^S(x)|^2 =
        P^S(\Omega)+eR .
    \end{aligned}
\end{equation}
Thus the absolute dipole depends on which cell is chosen to represent the periodic system. For the general many-particle expression, shifting the integration cell of the $j$th coordinate by a lattice vector $\bm{R}_j$ gives
\begin{equation}\label{eq:SI_branch_realspace}
    \bm{P}^{S}
    \rightarrow
    \bm{P}^{S}
    +
    e\sum_{j=1}^{N}(-1)^{\zeta_j}\bm{R}_j,
\end{equation}
because each coordinate-resolved probability integrates to unity. 

The same ambiguity appears in the Bloch representation as a change of home unit cell for the single-particle orbitals. If the orbital of band $n$ is reassigned by a lattice vector $\bm{R}_n$, then the cell-periodic Bloch function is replaced by
\(
    u_{n\sigma,\bm{k}}(\bm{r})
    \rightarrow
    e^{-i\bm{k}\cdot\bm{R}_n}
    u_{n\sigma,\bm{k}}(\bm{r}),
\)
which leaves the physical Bloch state single-valued because $e^{-i(\bm{k}+\bm{G})\cdot\bm{R}_n}=e^{-i\bm{k}\cdot\bm{R}_n}$ for any reciprocal-lattice vector $\bm{G}$, but corresponds to a phase $\chi_{n\sigma,\bm{k}}=\bm{k}\cdot\bm{R}_n$ with nontrivial winding on the Brillouin-zone torus. The diagonal Berry connection therefore shifts as
\(
    \bm{r}_{nn,\sigma,\bm{k}}
    \rightarrow
    \bm{r}_{nn,\sigma,\bm{k}}+\bm{R}_n,
\)
which is the reciprocal-space counterpart of Eq.~\eqref{eq:SI_branch_realspace}. 

Consequently, Eq.~\eqref{eq:SI_dipole_general} indicates that shift vector $\bm{X}^S_1$ is associated with the dipole up to a branch choice. Specifically, for neutral states with equal number of electrons and holes, one can fix the branch by choosing a consistent unit-cell convention (e.g. the center-of-mass reference) so that there is no relative reassignment of the electron and hole unit cells, making the branch dependence given in Eq.~\eqref{eq:SI_dipole_general} vanish. In experimental tests, one usually makes a unit-cell choice by exciting electrons at a specific real-space or momentum-space location, which also fixes the branch choice.


\section{Electron-Hole Sector and Real-Space Probability Density}\label{sec:SI_exciton}

\subsection{Electron-hole state and two-band reduction}\label{sec:SI_exciton_state}

For the electron-hole sector we absorb the spin index into the band label for notational simplicity. A correlated state with one electron and one hole and total momentum $\bm{Q}$ is
\begin{equation}\label{eq:SI_exciton_state}
    \ket{\Psi^{S,\bm{Q}}}
    =
    \sum_{cv\bm{k}}
    A^{S,\bm{Q}}_{cv\bm{k}}
    \ket{c,\bm{k}+\bm{Q}}_{\rm e}
    \ket{v,\bm{k}}_{\rm h},
\end{equation}
where $c$ and $v$ denote conduction- and valence-band indices. Its real-space wave function is
\begin{equation}\label{eq:SI_exciton_wavefunction}
    \begin{aligned}
        \Psi^{S}_{\bm{Q}}(\bm{r}_e,\bm{r}_h)
        &=
        \sum_{cv\bm{k}}
        A^{S,\bm{Q}}_{cv\bm{k}}
        e^{i\bm{Q}\cdot\bm{r}_e}
        e^{i\bm{k}\cdot(\bm{r}_e-\bm{r}_h)} \\
        &\quad\times
        u_{c,\bm{k}+\bm{Q}}(\bm{r}_e)
        u^*_{v,\bm{k}}(\bm{r}_h).
    \end{aligned}
\end{equation}

If the state is dominated by a single conduction band and a single valence band, Eq.\,\eqref{eq:SI_dipole_general} reduces to
\begin{equation}\label{eq:SI_X1_two_band}
    \bm{X}_1^{S,\bm{Q}}
    =
    \sum_{\bm{k}}
    \left|A^{S,\bm{Q}}_{cv\bm{k}}\right|^2
    \bm{R}^{S,\bm{Q}}_{cv,\bm{k}},
\end{equation}
with
\begin{equation}\label{eq:SI_R_S_Q}
    \bm{R}^{S,\bm{Q}}_{cv,\bm{k}}
    \equiv
    \nabla_{\bm{k}}\arg A^{S,\bm{Q}}_{cv\bm{k}}
    -
    \left(
        \bm{r}_{cc,\bm{k}+\bm{Q}}
        -
        \bm{r}_{vv,\bm{k}}
    \right).
\end{equation}
In the weak-hybridization limit, where one momentum $\bm{k}'$ dominates the profile function, $\bm{X}^{S,\bm{Q}}_1$ collapses to the corresponding individual shift vector $\bm{R}^{S,\bm{Q}}_{cv,\bm{k}'}$.

\subsection{Expansion of the probability density}\label{sec:SI_probability_expansion}

The joint electron-hole probability density is
\begin{equation}\label{eq:SI_probability_exact}
    \begin{aligned}
        \left|\Psi^{S}_{\bm{Q}}(\bm{r}_e,\bm{r}_h)\right|^2
        &=
        \sum_{cvc'v'}
        \sum_{\bm{k},\bm{q}}
        \left(A^{S,\bm{Q}}_{c'v',\bm{k}+\bm{q}}\right)^*
        A^{S,\bm{Q}}_{cv,\bm{k}}
        e^{-i\bm{q}\cdot(\bm{r}_e-\bm{r}_h)} \\
        &\quad\times
        u^*_{c',\bm{k}+\bm{q}+\bm{Q}}(\bm{r}_e)
        u_{c,\bm{k}+\bm{Q}}(\bm{r}_e)
        u_{v',\bm{k}+\bm{q}}(\bm{r}_h)
        u^*_{v,\bm{k}}(\bm{r}_h).
    \end{aligned}
\end{equation}
When the profile function is localized in momentum space, the dominant contributions come from small relative momenta $\bm{q}$. Expanding Eq.\,\eqref{eq:SI_probability_exact} to linear order in $\bm{q}$ gives
\begin{equation}\label{eq:SI_psi2_expansion}
    \begin{aligned}
        \left|\Psi^{S}_{\bm{Q}}(\bm{r}_e,\bm{r}_h)\right|^2
        &\approx
        Y_0(\bm{r}_e-\bm{r}_h)
        X^{S,\bm{Q}}_0(\bm{r}_e,\bm{r}_h) \\
        &\quad-
        \bm{Y}_1(\bm{r}_e-\bm{r}_h)\cdot
        \bm{X}^{S,\bm{Q}}_1(\bm{r}_e,\bm{r}_h)
        +
        \mathcal{O}(|\bm{q}|^2),
    \end{aligned}
\end{equation}
where
\begin{equation}\label{eq:SI_X0_X1_raw}
    \begin{aligned}
        X^{S,\bm{Q}}_0(\bm{r}_e,\bm{r}_h)
        &\equiv
        \sum_{\bm{k}}
        \sum_{cvc'v'}
        \rho_{c'c,\bm{k}+\bm{Q}}(\bm{r}_e)
        \rho_{vv',\bm{k}}(\bm{r}_h)
        \left(A^{S,\bm{Q}}_{c'v',\bm{k}}\right)^*
        A^{S,\bm{Q}}_{cv,\bm{k}}, \\
        \bm{X}^{S,\bm{Q}}_1(\bm{r}_e,\bm{r}_h)
        &\equiv
        i
        \sum_{\bm{k}}
        \sum_{cvc'v'}
        \rho_{c'c,\bm{k}+\bm{Q}}(\bm{r}_e)
        \rho_{vv',\bm{k}}(\bm{r}_h)
        \left(\nabla_{\bm{k}}A^{S,\bm{Q}}_{c'v',\bm{k}}\right)^*
        A^{S,\bm{Q}}_{cv,\bm{k}} \\
        &\quad+
        i
        \sum_{\bm{k}}
        \sum_{cvc'v'}
        \left(A^{S,\bm{Q}}_{c'v',\bm{k}}\right)^*
        A^{S,\bm{Q}}_{cv,\bm{k}}
        \rho_{vv',\bm{k}}(\bm{r}_h)
        \nabla_{\bm{k}}u^*_{c',\bm{k}+\bm{Q}}(\bm{r}_e)
        u_{c,\bm{k}+\bm{Q}}(\bm{r}_e) \\
        &\quad+
        i
        \sum_{\bm{k}}
        \sum_{cvc'v'}
        \left(A^{S,\bm{Q}}_{c'v',\bm{k}}\right)^*
        A^{S,\bm{Q}}_{cv,\bm{k}}
        \rho_{c'c,\bm{k}+\bm{Q}}(\bm{r}_e)
        \nabla_{\bm{k}}u_{v,\bm{k}}(\bm{r}_h)
        u^*_{v',\bm{k}}(\bm{r}_h).
    \end{aligned}
\end{equation}
The exciton-independent functions are
\begin{equation}\label{eq:SI_Y0_Y1_def}
    Y_0(\bm{r})
    \equiv
    \sum_{\bm{q}\in\mathcal{K}}e^{-i\bm{q}\cdot\bm{r}},
    \qquad
    \bm{Y}_1(\bm{r})
    \equiv
    i\sum_{\bm{q}\in\mathcal{K}}\bm{q}e^{-i\bm{q}\cdot\bm{r}},
\end{equation}
where $\mathcal{K}$ denotes the small momentum patch around the dominant exciton momenta.

Using Eq.\,\eqref{eq:SI_grad_cell_wavefunction}, Eq.\,\eqref{eq:SI_X0_X1_raw} can be rewritten as
\begin{equation}\label{eq:SI_X1_raw_rewritten}
    \begin{aligned}
        \bm{X}^{S,\bm{Q}}_1(\bm{r}_e,\bm{r}_h)
        &=
        i
        \sum_{\bm{k}}
        \sum_{cvc'v'}
        \rho_{c'c,\bm{k}+\bm{Q}}(\bm{r}_e)
        \rho_{vv',\bm{k}}(\bm{r}_h)
        \left(\nabla_{\bm{k}}A^{S,\bm{Q}}_{c'v',\bm{k}}\right)^*
        A^{S,\bm{Q}}_{cv,\bm{k}} \\
        &\quad-
        \sum_{\bm{k}}
        \sum_{cvc'v'}
        \sum_{m}
        \left(A^{S,\bm{Q}}_{c'v',\bm{k}}\right)^*
        A^{S,\bm{Q}}_{cv,\bm{k}}
        \bm{r}_{c'm,\bm{k}+\bm{Q}}
        \rho_{mc,\bm{k}+\bm{Q}}(\bm{r}_e)
        \rho_{vv',\bm{k}}(\bm{r}_h) \\
        &\quad+
        \sum_{\bm{k}}
        \sum_{cvc'v'}
        \sum_{m}
        \left(A^{S,\bm{Q}}_{c'v',\bm{k}}\right)^*
        A^{S,\bm{Q}}_{cv,\bm{k}}
        \bm{r}_{mv,\bm{k}}
        \rho_{v'm,\bm{k}}(\bm{r}_h)
        \rho_{c'c,\bm{k}+\bm{Q}}(\bm{r}_e).
    \end{aligned}
\end{equation}

\subsection{Coarse-grained lattice-resolved probability density}\label{sec:SI_probability_coarse}

To isolate the shift of the probability envelope, we average over intracell coordinates. Let
\begin{equation}
    \bm{r}_e=\bm{x}_i+\Delta\bm{r}_e,
    \qquad
    \bm{r}_h=\bm{x}_j+\Delta\bm{r}_h,
\end{equation}
where $\bm{x}_i$ and $\bm{x}_j$ are unit-cell centers and $\Delta\bm{r}_e$, $\Delta\bm{r}_h$ lie inside the unit cell $\Omega$. The scalar coefficient becomes
\begin{equation}
    \begin{aligned}
        X^{S,\bm{Q}}_0
        &\equiv
        \int_{\Omega}d^d\Delta\bm{r}_e
        \int_{\Omega}d^d\Delta\bm{r}_h\,
        X^{S,\bm{Q}}_0(\Delta\bm{r}_e,\Delta\bm{r}_h) \\
        &=
        \sum_{\bm{k}}
        \sum_{cv}
        \left|A^{S,\bm{Q}}_{cv,\bm{k}}\right|^2
        =1.
    \end{aligned}
\end{equation}
Applying the same coarse graining to Eq.\,\eqref{eq:SI_X1_raw_rewritten} gives
\begin{equation}\label{eq:SI_X1_macro}
    \begin{aligned}
        \bm{X}^{S,\bm{Q}}_1
        &=
        \int_{\Omega}d^d\Delta\bm{r}_e
        \int_{\Omega}d^d\Delta\bm{r}_h\,
        \bm{X}^{S,\bm{Q}}_1(\Delta\bm{r}_e,\Delta\bm{r}_h) \\
        &=
        i\sum_{\bm{k}}
        \sum_{cv}
        \left(\nabla_{\bm{k}}A^{S,\bm{Q}}_{cv,\bm{k}}\right)^*
        A^{S,\bm{Q}}_{cv,\bm{k}} \\
        &\quad-
        \sum_{\bm{k}}
        \sum_{cvc'}
        \left(A^{S,\bm{Q}}_{c'v,\bm{k}}\right)^*
        A^{S,\bm{Q}}_{cv,\bm{k}}
        \bm{r}_{c'c,\bm{k}+\bm{Q}} \\
        &\quad+
        \sum_{\bm{k}}
        \sum_{cvv'}
        \left(A^{S,\bm{Q}}_{cv',\bm{k}}\right)^*
        A^{S,\bm{Q}}_{cv,\bm{k}}
        \bm{r}_{v'v,\bm{k}}.
    \end{aligned}
\end{equation}
Equivalently,
\begin{equation}\label{eq:SI_X1_macro_compact}
    \bm{X}^{S,\bm{Q}}_1
    =
    \sum_{cv,\bm{k}}
    |A^{S,\bm{Q}}_{cv,\bm{k}}|^2
    \nabla_{\bm{k}}\arg A^{S,\bm{Q}}_{cv,\bm{k}}
    -
    \left\langle
        \Psi^{S,\bm{Q}}
        \middle|
        \hat{\bm{r}}
        \middle|
        \Psi^{S,\bm{Q}}
    \right\rangle,
\end{equation}
where the single-particle position operator is
\begin{equation}
    \hat{\bm{r}}
    \equiv
    \sum_{mn,\bm{k}}
    \bm{r}_{mn,\bm{k}}
    \hat{c}^{\dagger}_{m,\bm{k}}
    \hat{c}_{n,\bm{k}}.
\end{equation}
Its matrix element in the exciton basis is
\begin{equation}\label{eq:SI_exciton_r_matrix_element}
    \begin{aligned}
        &
        \left\langle
            \Psi^{S,\bm{Q}}
            \middle|
            \hat{\bm{r}}
            \middle|
            \Psi^{S,\bm{Q}}
        \right\rangle \\
        &=
        \sum_{mn,\bm{k}}
        \bm{r}_{mn,\bm{k}}
        \sum_{c_1v_1\bm{k}_1}
        \sum_{c_2v_2\bm{k}_2}
        \left(A^{S,\bm{Q}}_{c_1v_1,\bm{k}_1}\right)^*
        A^{S,\bm{Q}}_{c_2v_2,\bm{k}_2} \\
        &\quad\times
        \left\langle
            \hat{c}^{\dagger}_{v_1,\bm{k}_1}
            \hat{c}_{c_1,\bm{k}_1+\bm{Q}}
            \hat{c}^{\dagger}_{m,\bm{k}}
            \hat{c}_{n,\bm{k}}
            \hat{c}^{\dagger}_{c_2,\bm{k}_2+\bm{Q}}
            \hat{c}_{v_2,\bm{k}_2}
        \right\rangle \\
        &=
        \sum_{\bm{k}}
        \sum_{cv,c'v'}
        \left(A^{S,\bm{Q}}_{c'v',\bm{k}}\right)^*
        A^{S,\bm{Q}}_{cv,\bm{k}}
        \left(
            \bm{r}_{c'c,\bm{k}+\bm{Q}}\delta_{vv'}
            -
            \bm{r}_{v'v,\bm{k}}\delta_{cc'}
        \right).
    \end{aligned}
\end{equation}

The coarse-grained probability density then becomes
\begin{equation}\label{eq:SI_psi2_macro_localfield}
    \begin{aligned}
        \left|\Psi^{S}_{\bm{Q}}(\bm{x}_i-\bm{x}_j)\right|^2
        &\equiv
        \int_{\Omega}d^d\Delta\bm{r}_e
        \int_{\Omega}d^d\Delta\bm{r}_h\,
        \left|
            \Psi^{S}_{\bm{Q}}(\bm{x}_i+\Delta\bm{r}_e,\bm{x}_j+\Delta\bm{r}_h)
        \right|^2 \\
        &\approx
        Y_0(\bm{x}_i-\bm{x}_j)
        -
        \bm{Y}_1(\bm{x}_i-\bm{x}_j)\cdot
        \bm{X}^{S,\bm{Q}}_1.
    \end{aligned}
\end{equation}
For opposite separations $\bm{x}_i-\bm{x}_j=\pm Z\bm{a}_l$, with $\bm{a}_l$ a lattice vector and $Z$ an integer, Eq.\,\eqref{eq:SI_psi2_macro_localfield} gives
\begin{equation}\label{eq:SI_lattice_asymmetry}
    \bm{a}_l\cdot\bm{X}^{S,\bm{Q}}_1
    \propto
    \frac{
        \left|\Psi^S_{\bm{Q}}(Z\bm{a}_l)\right|^2
        -
        \left|\Psi^S_{\bm{Q}}(-Z\bm{a}_l)\right|^2
    }{Z}.
\end{equation}
This is the lattice-resolved inversion asymmetry quoted in the main text. If the exciton spans several well-separated momentum patches, additional interpatch terms oscillate on the lattice scale and can be separated from the coarse-grained envelope.

\subsection{Small-patch limit in two dimensions}\label{sec:SI_small_patch}

For low-energy excitons in a two-dimensional semiconductor, it is often sufficient to approximate the dominant momentum patch by a disk of radius $q_{\Lambda}$. In that case,
\begin{equation}\label{eq:SI_Y0_2D}
    \begin{aligned}
        Y_0(\bm{r})
        &=
        \int_{|\bm{q}|<q_{\Lambda}}
        \frac{d^2\bm{q}}{(2\pi)^2}
        e^{-i\bm{q}\cdot\bm{r}} \\
        &=
        \int_{0}^{q_{\Lambda}}qdq
        \int_{0}^{2\pi}\frac{d\varphi}{(2\pi)^2}
        e^{-iq|\bm{r}|\cos\varphi}
        =
        \frac{q_{\Lambda}^{2}}{2\pi}
        \frac{J_1(q_{\Lambda}|\bm{r}|)}{q_{\Lambda}|\bm{r}|},
    \end{aligned}
\end{equation}
and
\begin{equation}\label{eq:SI_Y1_2D}
    \begin{aligned}
        \bm{Y}_1(\bm{r})
        &=
        i
        \int_{|\bm{q}|<q_{\Lambda}}
        \frac{d^2\bm{q}}{(2\pi)^2}
        \bm{q}\,
        e^{-i\bm{q}\cdot\bm{r}} \\
        &=
        \frac{\bm{e}_{\bm{r}}}{2\pi}
        \int_{0}^{q_{\Lambda}}dq\,q^2J_1(q|\bm{r}|) \\
        &=
        \frac{q_{\Lambda}^{3}\bm{e}_{\bm{r}}}{2\pi}
        \frac{J_2(q_{\Lambda}|\bm{r}|)}{q_{\Lambda}|\bm{r}|},
    \end{aligned}
\end{equation}
where $J_n(x)$ is the Bessel function of the first kind and $\bm{e}_{\bm{r}}=\bm{r}/|\bm{r}|$. In the short-range regime $|\bm{r}|\ll q_{\Lambda}^{-1}$,
\begin{equation}\label{eq:SI_short_range_expansion}
    \begin{aligned}
        Y_0(\bm{r})
        &\approx
        \frac{q_{\Lambda}^2}{4\pi}
        \left[
            1-\frac{(q_{\Lambda}|\bm{r}|)^2}{8}
        \right], \\
        \bm{Y}_1(\bm{r})
        &\approx
        \bm{r}\,\frac{q_{\Lambda}^4}{16\pi}.
    \end{aligned}
\end{equation}
Substituting these expressions into Eq.\,\eqref{eq:SI_psi2_macro_localfield} gives
\begin{equation}\label{eq:SI_psi2_macro_localfield_2D}
    \begin{aligned}
        \left|\Psi^{S}_{\bm{Q}}(\bm{x}_i-\bm{x}_j)\right|^2
        &\propto
        1
        +
        \left(
            \frac{q_{\Lambda}|\bm{X}^{S,\bm{Q}}_1|}{2}
        \right)^2
        -
        \frac{q_{\Lambda}^2}{8}
        \left|
            \bm{x}_i-\bm{x}_j+\bm{X}^{S,\bm{Q}}_1
        \right|^2.
    \end{aligned}
\end{equation}
The peak of the coarse-grained probability density is therefore displaced from $\bm{x}_i-\bm{x}_j=\bm{0}$ to $\bm{x}_i-\bm{x}_j=-\bm{X}^{S,\bm{Q}}_1$. This interpretation is valid within the single-patch, short-range approximation used above.


\section{Optical Benchmark: Recovery of the Shift Current}\label{sec:SI_optical}

Even in the absence of intrinsic many-body interactions, a monochromatic optical field creates correlated electron-hole pairs. Consider the interaction-picture perturbation
\begin{equation}
    \begin{aligned}
        \hat{V}_I(t)
        &=
        e\,\hat{\bm{r}}(t)\cdot\bm{E}_0\cos(\Omega t)e^{-\gamma|t|} \\
        &=
        e|\bm{E}_0|\cos(\Omega t)e^{-\gamma|t|}
        \sum_{nm,\bm{k}}
        \bm{e}_p\cdot\bm{r}_{nm,\bm{k}}\,
        \hat{c}^{\dagger}_{n,\bm{k}}
        \hat{c}_{m,\bm{k}}
        e^{i\omega_{nm,\bm{k}}t},
    \end{aligned}
\end{equation}
where $\bm{e}_p=\bm{E}_0/|\bm{E}_0|$ and $\omega_{nm,\bm{k}}=(\epsilon_{n,\bm{k}}-\epsilon_{m,\bm{k}})/\hbar$. The interaction-picture evolution operator is the Dyson series reads
\begin{equation}
    \hat{U}_I(t,-\infty)
    =
    \mathcal{T}
    \exp\!\left[
        -\frac{i}{\hbar}
        \int_{-\infty}^{t}d\tau\,\hat{V}_I(\tau)
    \right].
\end{equation}
Keeping only the first-order term gives
\begin{equation}\label{eq:SI_psi0_t_optic}
    \begin{aligned}
        \ket{\Psi_0(t)}
        &\approx
        \ket{0}
        -
        \frac{i}{\hbar}
        \int_{-\infty}^{t}d\tau\,\hat{V}_I(\tau)\ket{0} \\
        &=
        \ket{0}
        -
        \frac{i e|\bm{E}_0|}{\hbar}
        \sum_{nm,\bm{k}}
        \bm{e}_p\cdot\bm{r}_{nm,\bm{k}}
        \ket{nm,\bm{k}} \\
        &\quad\times
        \left[
            \frac{1}{2i(\omega_{nm,\bm{k}}+\Omega-i\gamma)}
            +
            \frac{1}{2i(\omega_{nm,\bm{k}}-\Omega-i\gamma)}
        \right. \\
        &\qquad\left.
            +
            \frac{e^{i(\omega_{nm,\bm{k}}+\Omega+i\gamma)t}-1}
            {2i(\omega_{nm,\bm{k}}+\Omega+i\gamma)}
            +
            \frac{e^{i(\omega_{nm,\bm{k}}-\Omega+i\gamma)t}-1}
            {2i(\omega_{nm,\bm{k}}-\Omega+i\gamma)}
        \right],
    \end{aligned}
\end{equation}
where $\ket{nm,\bm{k}}\equiv \hat{c}^{\dagger}_{n,\bm{k}}\hat{c}_{m,\bm{k}}\ket{0}$. Focusing on the resonant set $C_{\Omega}$ satisfying $\omega_{cv,\bm{k}}=|\Omega|$, we obtain
\begin{equation}
    \ket{\Psi_0(t)}
    \rightarrow
    \ket{0}
    +
    \frac{i e|\bm{E}_0|}{\hbar}
    \left(
        \frac{2-e^{-\gamma t}}{2\gamma}
    \right)
    \sum_{cv\bm{k}\in C_{\Omega}}
    \bm{e}_p\cdot\bm{r}_{cv,\bm{k}}
    \ket{c,\bm{k}}_{\rm e}\ket{v,\bm{k}}_{\rm h}.
\end{equation}
The optical field therefore prepares an electron-hole correlated state with profile function proportional to $\bm{e}_p\cdot\bm{r}_{cv,\bm{k}}$. Neglecting band degeneracies and using Eq.\,\eqref{eq:SI_X1_two_band}, the induced intrinsic dipole is
\begin{equation}\label{eq:SI_optical_dipole}
    \begin{aligned}
        e\bm{X}^{\Omega,\bm{E}_0}_1
        &=
        \frac{e^3|\bm{E}_0|^2}{\hbar^2}
        \left(
            \frac{2-e^{-\gamma t}}{2\gamma}
        \right)^2 \\
        &\quad\times
        \sum_{cv\bm{k}\in C_{\Omega}}
        \left|
            \bm{e}_p\cdot\bm{r}_{cv,\bm{k}}
        \right|^2
        \bm{R}^{\bm{e}_p}_{cv,\bm{k}},
    \end{aligned}
\end{equation}
where the conventional optical shift vector is
\begin{equation}\label{eq:SI_R_optic}
    \bm{R}^{\bm{e}_p}_{cv,\bm{k}}
    =
    \bm{r}_{cc,\bm{k}}
    -
    \bm{r}_{vv,\bm{k}}
    -
    \nabla_{\bm{k}}
    \arg\!\left(
        \bm{e}_p\cdot\bm{r}_{cv,\bm{k}}
    \right).
\end{equation}
The current is the time derivative of the polarization,
\begin{equation}
    \bm{J}(t)
    =
    \frac{d}{dt}
    \left(
        e\bm{X}^{\Omega,\bm{E}_0}_1
    \right).
\end{equation}
In the steady-illumination limit $\gamma\rightarrow 0$, the leading contribution is the dc current
\begin{equation}\label{eq:SI_shift_current_dc}
    \begin{aligned}
        \lim_{\gamma\rightarrow 0}\bm{J}(t)
        &=
        -\frac{e^3|\bm{E}_0|^2}{2\hbar^2}
        \lim_{\gamma\rightarrow 0}
        \frac{1}{\gamma}
        \sum_{cv\bm{k}\in C_{\Omega}}
        \left|
            \bm{e}_p\cdot\bm{r}_{cv,\bm{k}}
        \right|^2
        \bm{R}^{\bm{e}_p}_{cv,\bm{k}} \\
        &\equiv
        \bm{J}_{\rm dc}(\Omega,\bm{E}_0).
    \end{aligned}
\end{equation}
The corresponding conductivity is
\begin{equation}\label{eq:SI_shift_current_conductivity}
    \begin{aligned}
        \sigma_{\mu\bm{e}_p}(\Omega)
        &\equiv
        \frac{\partial^2 J^{\mu}_{\rm dc}(\Omega,\bm{E}_0)}
        {\partial |\bm{E}_0|^2} \\
        &=
        -\frac{\pi e^3}{\hbar^2}
        \sum_{cv\bm{k}}
        \left|
            \bm{e}_p\cdot\bm{r}_{cv,\bm{k}}
        \right|^2
        R^{\mu\bm{e}_p}_{cv,\bm{k}}
        \delta(\omega_{cv,\bm{k}}-|\Omega|),
    \end{aligned}
\end{equation}
where we used the identity
\( \lim_{\gamma \rightarrow 0} \frac{1}{\gamma} \sum_{cv\bm{k}\in C_\Omega} = \sum_{cv\bm{k}}2\pi\delta(\omega_{cv\bm{k}}-\Omega) \). It reproduces the standard shift-current result~\cite{cook2017design,sipe2000second,young2012first}. This benchmark shows explicitly that the phase-gradient term in the optical shift vector is the coherence contribution to the intrinsic dipole of the optically generated electron-hole correlation.


\section{Two-Electron Sector: Cooper-Pair Mapping}\label{sec:SI_cooper}

The same logic can be applied to a two-electron correlated state. Consider a Cooper-pair basis built from states with opposite momenta and opposite spins,
\begin{equation}\label{eq:SI_cooper_basis}
    \ket{c_1\sigma,\bm{k}}_{\rm e}
    \ket{c_2\bar{\sigma},-\bm{k}}_{\rm e},
\end{equation}
where $c_1$ and $c_2$ label conduction-band indices. Its real-space wave function reads
\begin{equation}\label{eq:SI_cooper_realspace_basis}
    \langle \bm{r}_1 \mid c_1\sigma,\bm{k}\rangle_{\rm e}
    \langle \bm{r}_2 \mid c_2\bar{\sigma},-\bm{k}\rangle_{\rm e}
    =
    e^{i\bm{k}\cdot(\bm{r}_1-\bm{r}_2)}
    u_{c_1\sigma,\bm{k}}(\bm{r}_1)
    u_{c_2\bar{\sigma},-\bm{k}}(\bm{r}_2).
\end{equation}
In a time-reversal-symmetric system,
\begin{equation}
    \mathcal{T}\ket{c\sigma,\bm{k}}_{\rm e}
    =
    e^{i\chi_{c\sigma,\bm{k}}}
    \ket{c\bar{\sigma},-\bm{k}}_{\rm e},
\end{equation}
which implies
\begin{equation}\label{eq:SI_TR_hole_mapping}
    \langle \bm{r}\mid c\bar{\sigma},-\bm{k}\rangle_{\rm e}
    =
    e^{i\chi_{c\sigma,\bm{k}}}
    \langle \bm{r}\mid c\sigma,\bm{k}\rangle_{\rm h}.
\end{equation}
Equation~\eqref{eq:SI_TR_hole_mapping} shows that the second electron can be viewed as an effective hole degree of freedom, up to an arbitrary phase. A general Cooper-pair state can then be written as
\begin{equation}
    \ket{\Psi^{\lambda}}
    =
    \sum_{c_1c_2\bm{k}}
    B^{\lambda}_{c_1c_2\bm{k}}
    \ket{c_1\sigma,\bm{k}}_{\rm e}
    \ket{c_2\bar{\sigma},-\bm{k}}_{\rm e}
\end{equation}
or, equivalently,
\begin{equation}\label{eq:SI_cooper_effective_eh}
    \ket{\Psi^{\lambda}}
    =
    \sum_{c_1c_2\bm{k}}
    \widetilde{A}^{\lambda}_{c_1c_2\bm{k}}
    \ket{c_1,\bm{k}}_{\rm e}
    \ket{c_2,\bm{k}}_{\rm h},
    \qquad
    \widetilde{A}^{\lambda}_{c_1c_2\bm{k}}
    \equiv
    e^{i\chi_{c_2\sigma,\bm{k}}}
    B^{\lambda}_{c_1c_2\bm{k}}.
\end{equation}
The Cooper-pair problem therefore has the same formal structure as the electron-hole problem with zero total momentum, except that the effective hole index runs inside the conduction manifold. The generalized shift vector constructed from $\widetilde{A}^{\lambda}_{c_1c_2\bm{k}}$ thus measures the internal relative-coordinate shift of the two-electron state.



\section{First-principles exciton calculation}\label{sec:first_principles_MoS2}

We calculate the \textit{ab initio} excitonic properties of monolayer MoS$_2$ by solving the Bethe-Salpeter equation (BSE)~\cite{rohlfing2000electron} as implemented in the {\sc BerkeleyGW} package~\cite{deslippe2012berkeleygw}. We obtain the underlying single-particle electronic structure from density functional theory calculations performed using {\sc Quantum ESPRESSO}~\cite{giannozzi2009quantum}, which provide the electron and hole wave functions used in the BSE. We employ optimized norm-conserving Vanderbilt pseudopotentials~\cite{hamann2013optimized}, and treat the exchange-correlation functional within the generalized gradient approximation of Perdew, Burke, and Ernzerhof~\cite{perdew1996generalized}. We include spin-orbit coupling explicitly. We evaluate the self-consistent charge density on a $12\times 12\times 1$ Monkhorst-Pack $\bm{k}$-point grid~\cite{monkhorst1976special}. We use a lattice constant of $3.15$~\AA, together with a vacuum spacing of $15$~\AA\@ to suppress spurious interactions between periodic images.

We evaluate the dielectric screening on a $48\times 48\times 1$ $\bm{k}$-point grid. To remove artificial interactions between repeated slabs, we employed a truncation scheme for the Coulomb interaction along the out-of-plane direction~\cite{ismail2006truncation}. We construct the BSE Hamiltonian using two occupied and two unoccupied spinor states. We use a $48\times 48\times 1$ coarse Brillouin-zone sampling and then interpolate onto a finer $96\times 96\times 1$ grid to evaluate the electron-hole interaction kernel and obtain converged excitonic states.

\end{document}